\documentclass[aps,amsfonts,pra,superscriptaddress,twocolumn,showpacs]{revtex4}
\usepackage{epsfig,amsmath,amssymb,bm,epsf,graphicx,psfrag,float}
\usepackage{color}

\def\bra#1{\langle#1\vert}
\def\ket#1{\vert#1\rangle}
\def\ketbra#1{\vert#1\rangle\langle#1\vert}

\def\Longarrow{\protect\@lra}
\def\@lra{\relbar\joinrel\relbar\joinrel\relbar\joinrel%
          \relbar\joinrel\rightarrow}

\newcommand{\bc}{\begin{center}}
\newcommand{\ec}{\end{center}}
\newcommand{\be}{\begin{equation}}
\newcommand{\ee}{\end{equation}}
\newcommand{\bea}{\begin{eqnarray}}
\newcommand{\eea}{\end{eqnarray}}

\newcommand{\ncd}{\newcommand}
\ncd{\QCcns}{$QC_{\cal{C}}$}
\ncd{\QCc}{$QC_{\cal{C}}\;$}

\definecolor{libl}{cmyk}{0.2,0.1,0,0}

\begin{document}

\title{Quantum computational universality of the Cai-Miyake-D\"ur-Briegel 2D
quantum state from Affleck-Kennedy-Lieb-Tasaki quasichains}

\author{Tzu-Chieh Wei}
\affiliation{Department of Physics and Astronomy, University of
British Columbia, Vancouver, British Columbia V6T 1Z1, Canada}
\affiliation{C. N. Yang Institute for Theoretical Physics, State
University of New York at Stony Brook, Stony Brook, NY 11794-3840,
USA}
\author{Robert Raussendorf}
\affiliation{Department of Physics and Astronomy, University of
British Columbia, Vancouver, British Columbia V6T 1Z1, Canada}
\author{Leong Chuan Kwek} \affiliation{Centre for Quantum
 Technologies, National University of Singapore, 2 Science Drive 3,
Singapore and National Institute of Education and Institute of
 Advanced Studies, Nanyang Technological University, 1 Nanyang Walk,
 Singapore
}
\date{\today}

\begin{abstract}
Universal quantum computation can be achieved by simply performing
single-qubit measurements on a highly entangled resource state, such
as cluster states. Cai, Miyake, D\"ur, and Briegel recently
constructed a ground state of a two-dimensional quantum magnet by
combining multiple Affleck-Kennedy-Lieb-Tasaki quasichains of mixed
spin-3/2 and spin-1/2 entities and by mapping pairs of neighboring
spin-1/2 particles to individual spin-3/2 particles [Phys. Rev. A
{\bf 82}, 052309 (2010)]. They showed that this state enables
universal quantum computation by single-spin measurements. Here, we
give an alternative understanding of how this state gives rise to
universal measurement-based quantum computation: by local
operations, each quasichain can be converted to a 1D cluster state
and entangling gates between two neighboring logical qubits can be
implemented by single-spin measurements. We further argue that a 2D
cluster state can be distilled from the Cai-Miyake-D\"ur-Briegel
state.
\end{abstract}
 \maketitle

 \section{ Introduction} Measurement-based quantum computation
(MBQC) provides a framework where quantum
computation~\cite{NielsenChuang00} is achieved by single-spin
measurements on a highly entangled resource
state~\cite{Oneway,Oneway2}. The first known resource state is the
so-called cluster state on a square lattice~\cite{Cluster}. Graph
states~\cite{Hein} on other two-dimensional regular lattices were
also shown to be universal resources~\cite{Universal}. It turns out
that universal resource states are shown to be rare, of zero
measure~\cite{Gross1}. However, resource states beyond graph states
have been constructed~\cite{Gross}, mostly due to the understanding
of MBQC from the perspective of matrix-product-states (MPS) and
projected-entangled-pair-states
(PEPS)~\cite{Verstraete,correlation}. In fact, there exist universal
resource states that are unique ground states of two-body
interacting
Hamiltonians~\cite{BartlettRudolph,Chen,Cai10,Honeycomb,Honeycomb2},
even though cluster states are not~\cite{Nielsen}.

Among the small class of known universal states is the construction
by Cai, Miyake, D\"ur, and Briegel (CMDB)~\cite{Cai10}. It is based
on chains of Affleck-Kennedy-Lieb-Tasaki (AKLT) states~\cite{AKLT},
originally constructed in the setting of condensed matter physics.
It was shown earlier that one-dimensional spin-1 AKLT states can
serve as resources for restricted quantum
computations~\cite{Gross,Brennen}, and universal computation can be
achieved by {\it active} coupling of many such 1D
chains~\cite{Brennen}. The merit of the CMDB
construction~\cite{Cai10} is to avoid such active coupling by
mapping two connecting spin-1/2 entities into one spin-3/2 entity,
and patching many 1D AKLT quasichains into a two-dimensional
structure. Moreover, the gap above the ground state of the
constructed 2D model was shown to be finite~\cite{Cai10}. Here, we
show that by local operations, the ground state of an AKLT
quasichain of mixed spin-3/2 and spin-1/2 entities can be converted
to a 1D cluster state. Furthermore, entangling gates between two
logical qubits on neighboring quasichains can be implemented by
measuring the spin-3/2 particle connecting them. This provides an
alternative understanding of the quantum-computational universality
using the CMDB state. Moreover, we argue that from this state a 2D
cluster state can be distilled by local operations.

The structure of the remaining of the paper is as follows. In
Sec.~\ref{sec:1Dquasichain} we review the 1D AKLT quasichain. In
Sec.~\ref{sec:2D} we review the 2D Cai-Miyake-D\"ur-Briegel model.
In Sec.~\ref{sec:1Dcluster} we show how the 1D quasichain AKLT state
can be locally converted to a 1D cluster state. In
Sec.~\ref{sec:2Dcluster} we show how to understand the quantum
computational universality of the CMDB state and how it can be
locally converted to a 2D cluster state. In
Sec.~\ref{sec:conclusion} we make some concluding remarks.

\section{1D AKLT quasichain}\label{sec:1Dquasichain}
\begin{figure}
  \begin{center}
 \includegraphics[width=8cm]{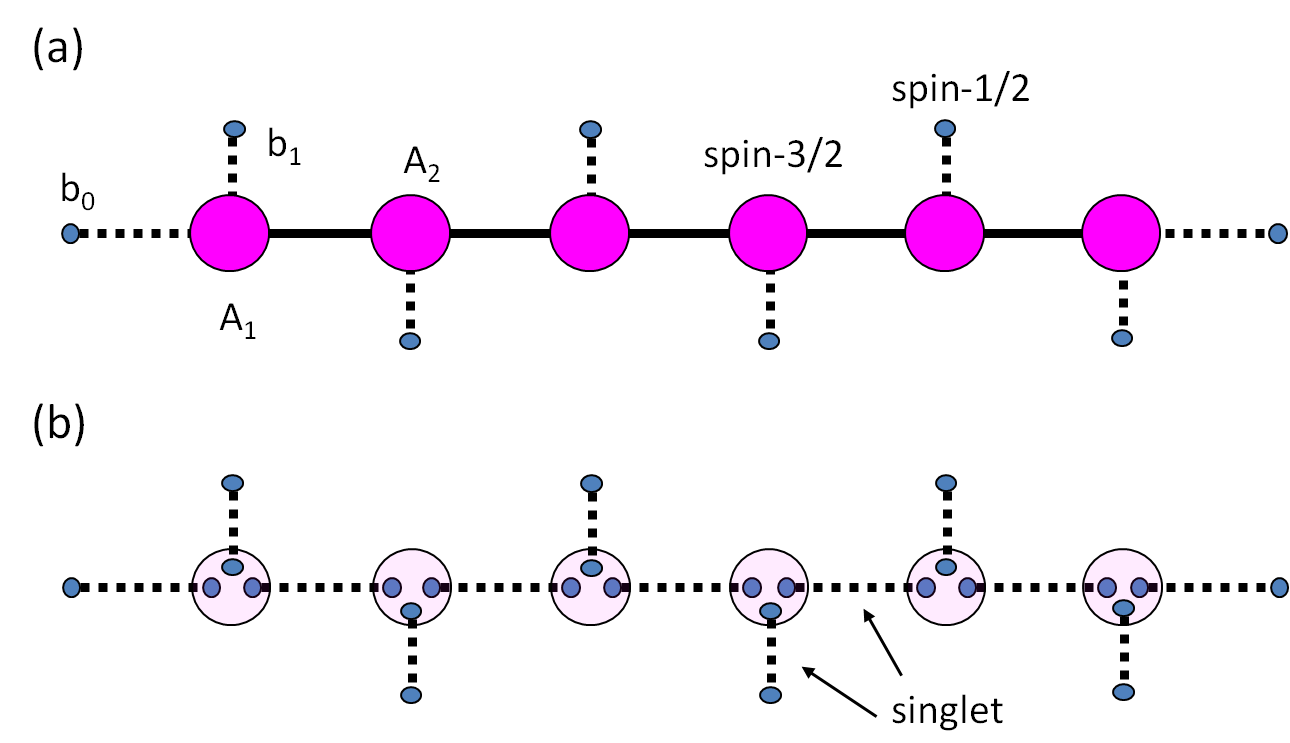}
    \end{center}
  \caption{\label{fig:fig1D}(Color
  online) 1D AKLT quasichain. (a) The larger circles represent spin-3/2 particles and are connected by thick solid lines. They form the backbone of the quasichain. The smaller
  circles represent spin-1/2 particles, which are connected via dashed lines to corresponding spin-3/2 particles.
  (b) The schematic of the ground state of the 1D AKLT quasichain. Each spin-3/2 can be regarded as
  three virtual spin-1/2 particles, each forming a singlet with the neighbor virtual qubit. The
  three qubits at $A$ site is then projected into their symmetric subspace, with the projection indicated by a transparent
  circle. }
\end{figure}
In~\cite{Cai10} the 1D AKLT model defined on a quasichain is
considered, which consists of spin-3/2 particles (labeled by
$A_i$'s) located at the backbone and spin-1/2 particles (labeled by
$b_i$'s) connected to $A$'s, as shown in Fig.~\ref{fig:fig1D}a. The
Hamiltonian of the quasichain is~\cite{Cai10}
\begin{equation}
\label{eqn:H1D}
 H=\sum_{i=1}^{N-1} P^{S=3}_{A_i,A_{i+1}}
+\sum_{i=1}^N P^{S=2}_{A_i,b_i}
+P^{S=2}_{A_1,b_0}+P^{S=2}_{A_N,b_{N+1}},
\end{equation}
where $P^{S}_{u,v}$ represents the projector onto the  spin-$S$
subspace of the joint $u$ and $v$ spins, which can be expressed as a
polynomial of $\vec{S}_u\cdot\vec{S}_v$. The original AKLT chain is
defined on a chain of spin-1 particles, with the ends terminated by
spin-1/2 particles~\cite{AKLT}. The quasichain, similar to the
original AKLT chain, can be shown to possess a unique ground state
with a finite spectral gap~\cite{Cai10}. As illustrated in
 Fig.~\ref{fig:fig1D}b, the ground state is a valence-bond-solid
state of the AKLT type, in which every $A$ is composed of three
virtual spin-1/2 particles, each forming a singlet pair with its
neighboring spin-1/2 particle, followed by a projection $\Pi_S$
(defined below) of the three virtual qubits to their symmetric
subspace. To be more precise, the ground state is the following
valence-bond state,
\begin{equation}
\label{eqn:vbs}
 |\psi_{\rm AKLT}\rangle\sim\mathop{\otimes}_{A}
\Pi_{S,A} \mathop{\otimes}_{{\rm edge}\, e} |\phi\rangle_e,
\end{equation}
where $|\phi\rangle_e=(|01\rangle-|10\rangle)/\sqrt{2}$ is a singlet
state for the pair of qubits residing on the (dashed) edge $e$; see
Fig.~\ref{fig:fig1D}b.

 The correspondence of the states of the three
virtual particles to those of a spin-3/2 particle is given as
follows,
\begin{subequations}
\label{eqn:relabel}
\begin{eqnarray}
\!\!\!\!\!\!\!\!\!\!\!\!&&\ket{000}\leftrightarrow \Big\vert\frac{3}{2},\frac{3}{2}\Big\rangle,\\
\!\!\!\!\!\!\!\!\!\!\!\!&&\ket{111}\leftrightarrow \Big\vert\frac{3}{2},-\frac{3}{2}\Big\rangle,\\
\!\!\!\!\!\!\!\!\!\!\!\!&&\ket{W}\equiv\frac{1}{\sqrt{3}}(\ket{001}+\ket{010}+\ket{100})\leftrightarrow
\Big\vert\frac{3}{2},\frac{1}{2}\Big\rangle,\\
\!\!\!\!\!\!\!\!\!\!\!\!&&\ket{\overline{W}}\equiv\frac{1}{\sqrt{3}}(\ket{110}+\ket{101}+\ket{011})\leftrightarrow
\Big\vert\frac{3}{2},-\frac{1}{2}\Big\rangle,
\end{eqnarray}
\end{subequations}
where the qubit states $|0/1\rangle$ are the eigenstates of spin-1/2
angular momentum operators $\hat{S}^2$ and $\hat{S}_z$, i.e.,
$|0\rangle\equiv |1/2,1/2\rangle$ and $|1\rangle\equiv
|1/2,-1/2\rangle$. The projector onto the symmetric subspace can be
simply written as $\Pi_S\equiv |000\rangle\langle 000|+  |W\rangle
\langle W| +|\overline{W}\rangle \langle \overline{W}| + |111
\rangle\langle 111|$. The quasichain AKLT state in
Eq.~(\ref{eqn:vbs}) is written in terms of the virtual-qubit
representation for the central spin-3/2 particles; the relabeling in
Eq.~(\ref{eqn:relabel}) provides a translation to the spin-3/2
representation. We note that the quantization-axis label on the
states is implicitly assumed to be along the $z$-axis. In the
following, when necessary, we will explicitly write out the
quantization axis.

\section{2D Cai-Miyake-D\"ur-Briegel model}\label{sec:2D}
\begin{figure}
   \includegraphics[width=8cm]{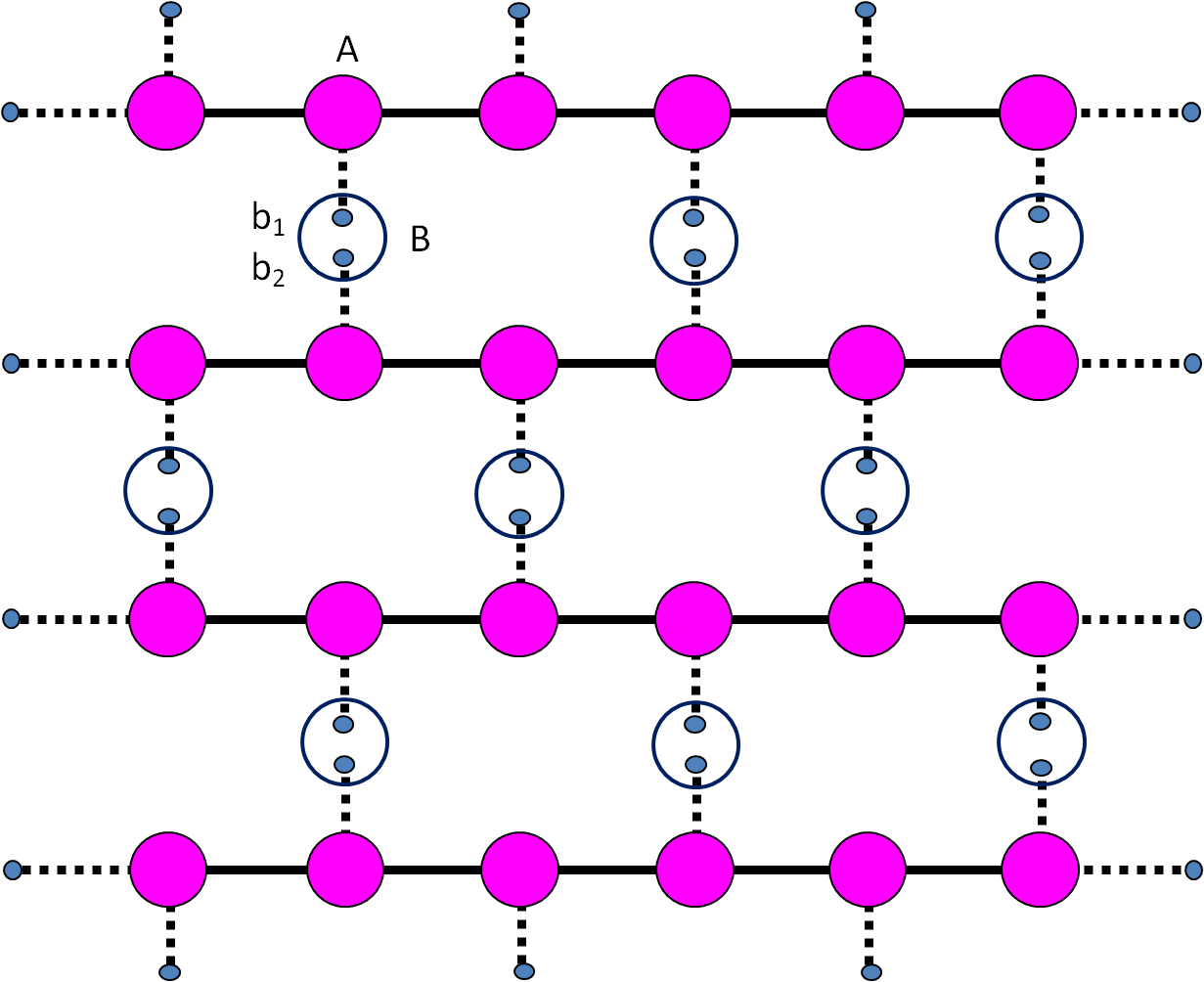}
  \caption{\label{fig:fig2D-CMDB}(Color
  online) The 2D Cai-Miyake-D\"ur-Brigel model. The model consists of many AKLT quasichains and
  merge the two neighboring spin-1/2 particles, e.g., $b_1$ and $b_2$, to a single spin-3/2 particle
  $B$ via $U$ in Eq.~(\ref{eqn:U}).}
\end{figure}
To build up a 2D model with a finite gap above the ground state, Cai
et al. stack up  1D quasichains and their  mirror images in a
staggered way, as shown in Fig.~\ref{fig:fig2D-CMDB}.  The two
neighboring spin-1/2 particles, e.g., $b_1$ and $b_2$, possess in
total four levels and can be formally mapped to a spin-3/2 particle
$B$ by a unitary transformation $U$,
\begin{equation}\label{eqn:U}
 U=\sum_{m_1,m_2=\pm 1/2} \Big\vert \frac{3}{2},m_1+2m_2\Big\rangle_{B}\Big\langle
  \frac{1}{2},m_1\Big\vert_{b_1}\Big\langle\frac{1}{2},m_2\Big\vert_{b_2}. \end{equation}
The Hamiltonian of the resulting 2D system is obtained from the sum
of Hamiltonians of many 1D quasichains~(\ref{eqn:H1D}), followed by
the unitary transformations $U$'s merging all pairs of neighboring
$b$'s to single spin-3/2 $B$'s. For example, a term
$f(\vec{S}_A\cdot \vec{s}_{b_1})$ in the Hamiltonian~(\ref{eqn:H1D})
is transformed to $f(\vec{S}_A\cdot U\vec{s}_{b_1}U^\dagger)$, with
$U\vec{s}_{b_1} U^\dagger$ and similarly  $U\vec{s}_{b_2} U^\dagger$
being observables of the spin-3/2 $B$. For the explicit form of the
resulting 2D Hamiltonian, we refer the readers to the paper by Cai
et al.~\cite{Cai10}. Because of the spectral gap in the quasichains,
the whole system by patching them together using the unitary $U$'s
has a gap as well. We note that in the following discussions of
quantum computational universality it is actually convenient to
treat $B$ in terms of the two virtual spin-1/2 $b$ particles before
the unitary $U$. Any unitary transformation on the composite $b_1$
and $b_1$ is a local unitary transformation on $B$. Moreover, joint
 measurements on $b_1$ and $b_2$ is a local measurement on $B$.
We also remark that with open boundary conditions the CMDB model
contains
 boundary spin-1/2 particles. To remove these and make the model
 consist solely of spin-3/2 particles, one can use periodic boundary
 conditions by merging two associated opposite-end spin-1/2's to one spin-3/2.

\section{From the ground state of the AKLT quasichain to a 1D cluster
state}\label{sec:1Dcluster} In this section we consider the ground
state of the AKLT quasichain and shall show that it can be locally
converted to a 1D cluster state. The local conversion is achieved by
a generalized measurement, also called positive-operator-value
measure (POVM)~\cite{NielsenChuang00} at each site, as was used
previously in the 2D AKLT state on the honeycomb
lattice~\cite{Honeycomb}. The idea is to ``project'' the four levels
of a spin-3/2 particle down to two levels, forming an effective
qubit. To bring out the ``hidden'' cluster state from the quasichain
AKLT state, we generalize the projection and introduce the following
POVM at each $A$ site, which consists of three rank-two
elements~\cite{Honeycomb}
\begin{subequations}
\label{POVM1}
  \begin{eqnarray}
\!\!\!\!\!\!\!\!\!\!{F}_{A,z}&=&\sqrt{\frac{2}{3}}(\ketbra{+3/2}_z+\ketbra{-3/2}_z), \\
\!\!\!\!\!\!\!\!\!\!{F}_{A,x}&=&\sqrt{\frac{2}{3}}(\ketbra{+3/2}_x+\ketbra{-3/2}_x),\\
\!\!\!\!\!\!\!\!\!\!{F}_{A,y}&=&\sqrt{\frac{2}{3}}(\ketbra{+3/2}_y+\ketbra{-3/2}_y),
\end{eqnarray}
\end{subequations}
where the subscripts $x$, $y$ and $z$ indicating the measurement
outcome in fact represent the effective quantization axes and we
have omitted the total spin magnitude $S=3/2$ in the Dirac brackets.
In contrast to the usual orthogonal or von Neumann measurement, the
POVM elements need not be orthogonal, as $F_{A,x}F_{A,z}\ne 0$. But
similar to von Neumann measurement, given the outcome $a$ of the
POVM $a=x$, $y$ or $z$, the state $|\psi\rangle$ undergoing the
measurement becomes $|\psi\rangle \rightarrow
F_{A,a}|\psi\rangle$~\cite{NielsenChuang00}. It can be verified that
these POVM elements satisfy the conservation of probability, i.e.,
$\sum_{\nu \in \{x,y,z\}}F^\dagger_{A,\nu} F_{A,\nu} = I_{S=3/2}$,
where $I_{S=3/2}$ is the identity operator for the spin-3/2 Hilbert
space. Physically, $F_{A,\nu}=({S}_{A,\nu}^2-1/4)/\sqrt{6}$ is
proportional to a projector onto the two-dimensional subspace
spanned by the two eigenstates of the $\nu$-component spin operator
with eigenvalues $S_\nu=\pm 3/2$. From the viewpoint of three
virtual qubits, the POVM elements can be written as
\begin{subequations}
\label{POVM2}
  \begin{eqnarray}
\!\!\!\!\!\!\!\!\!\!\tilde{F}_{A,z}&=&\sqrt{\frac{2}{3}}(\ketbra{000}+\ketbra{111}), \\
\!\!\!\!\!\!\!\!\!\!\tilde{F}_{A,x}&=&\sqrt{\frac{2}{3}}(\ketbra{+++}+\ketbra{---}),\\
\!\!\!\!\!\!\!\!\!\!\tilde{F}_{A,y}&=&\sqrt{\frac{2}{3}}(\ketbra{i,i,i}+\ketbra{-\!i,-\!i,-\!i}),
\end{eqnarray}
\end{subequations}
where $|0/1\rangle$, $\ket{\pm}\equiv(\ket{0}\pm\ket{1})/\sqrt{2}$
and $\ket{\pm i}\equiv (\ket{0}\pm i\ket{1})/\sqrt{2}$ are
eigenstates of Pauli operators $\sigma_z$, $\sigma_x$ and
$\sigma_y$, respectively. Note for convenience we shall also use
$\ket{0/1}_a$ (with $a=x$, $y$, or $z$) to denote eigenstates of
$\sigma_a$, i.e., $\sigma_a\ket{0/1}_a=\pm\ket{0/1}_a$. We note that
in this notation, $\tilde{F}$'s can be conveniently written as
\begin{equation}
\label{POVM3}
\tilde{F}_{A,a}=\sqrt{\frac{2}{3}}(\ket{000}_a\bra{000}+\ket{111}_a\bra{111}).
\end{equation}
One advantage of using the virtual-qubit representation is that the
stabilizer formalism~\cite{Stabilizer} can be employed and insight
about the post-POVM state can thus be gained. The above POVM
elements in the second form obey the relation $\sum_{\nu \in
\{x,y,z\}}\tilde{F}^\dagger_{A,\nu} \tilde{F}_{A,\nu} = \Pi_{S,A}$,
i.e., project onto the symmetric subspace, as required. The outcome
of the POVM at any site $A$ is random, which can be $x$, $y$ or $z$,
and it can be correlated with the outcomes at other sites due to the
entanglement in the AKLT state~\cite{AKLT}.

One important consequence after the POVM on a spin-3/2 $A$ is that,
even though it has four levels, its state after the POVM is
restricted to the two-dimensional Hilbert subspace, spanned by
$|+3/2\rangle$ and $|-3/2\rangle$, or in terms of the three virtual
qubits by $|000\rangle$ and $|111\rangle$ with the quantization axis
given by the POVM outcome. Thus, after the POVM, one can treat each
$A$ site as the carrier of one qubit.

\subsection{Strategy}
Given the outcomes $\{a_v\}$ of POVMs at all center sites $\{v\}$,
we know the post-POVM state is given by
\begin{subequations}
\begin{eqnarray}
|\psi'\rangle &\sim& \mathop{\otimes}_{v} \tilde{F}_{v,a_v}
|\psi_{\rm AKLT}\rangle\\
&\sim& \mathop{\otimes}_{v} \tilde{F}_{v,a_v} \mathop{\otimes}_{{\rm
edge}\, e} |\phi\rangle_e,
\end{eqnarray}
\end{subequations}
where we have used the virtual-qubit version of $F$'s and in going
from the first line to the second, we have used the fact that
$\tilde{F}$'s projects into the symmetric subspace of three qubits
as well and thus there is no need to keep the projectors
$\Pi_{S,v}$'s. To understand what the post-POVM state is, we then
employ the stabilizer formalism~\cite{Stabilizer} and try to find
its  stabilizer operators. For a singlet state
$|\phi\rangle_{12}=(|01\rangle-|10\rangle)/\sqrt{2}$ of two qubits
$1$ and $2$, it is easy to see that there are stabilizer
operators~\cite{stabilizer2} $-\sigma_a^{[1]}\otimes\sigma_a^{[2]}$
with $a=x$, $y$, or $z$, namely,
\begin{equation}
-\sigma_x\otimes\sigma_x|\phi\rangle=-\sigma_y\otimes\sigma_y|\phi\rangle=-\sigma_z\otimes\sigma_z|\phi\rangle
=|\phi\rangle.
\end{equation}
The key ingredient of the stabilizer formalism applied to this
example is that given two of the above commuting and independent
operators, e.g., $-\sigma_x\otimes\sigma_x$ and
$-\sigma_z\otimes\sigma_z$, the state $|\phi\rangle$ is uniquely
determined. If operators such as the above commute with all
$\tilde{F}$'s, they will remain the stabilizer operators for the
post-POVM state.

Another type of stabilizer operators arises from the POVMs. For
example, using the form given in Eq.~(\ref{POVM3}), we see that
there are operators (two of them being independent) that satisfy
\begin{equation}
\label{eqn:stabilizerF} \sigma_a^{[1]}\otimes \sigma_a^{[2]}
\tilde{F}_{A,a} =\sigma_a^{[1]}\otimes \sigma_a^{[3]}
\tilde{F}_{A,a} =\sigma_a^{[2]}\otimes \sigma_a^{[3]}
\tilde{F}_{A,a}= \tilde{F}_{A,a}.
\end{equation}
Hence, these are stabilizer operators of the post-POVM state.

A third type of stabilizer operators that we shall identify below
give rise to the graph-state stabilizer operators. These emerge from
POVMs across a few sites. Once we identify all stabilizer operators,
the state is uniquely determined if the number of independent
stabilizer operators equals that of spins.
\subsection{Encoding}
\label{sec:encoding}
\begin{figure}
\begin{center}
  \includegraphics[width=8cm]{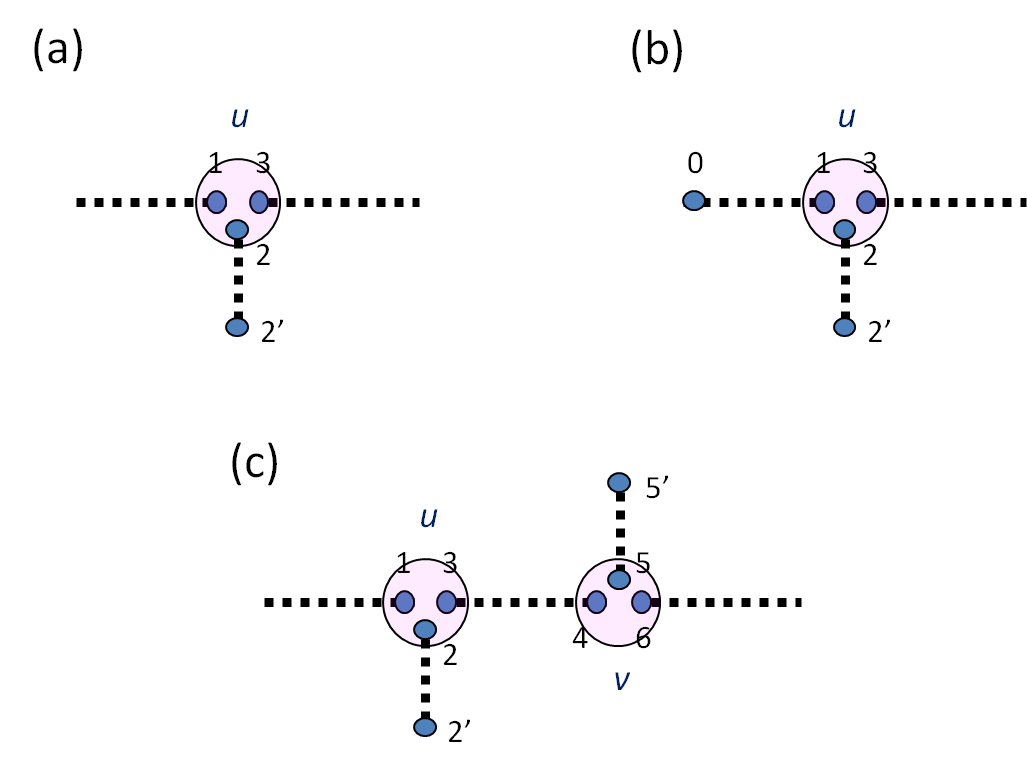}
\end{center}
  \caption{\label{fig:figEncoding}(Color
  online) Encoding of a qubit. (a) The encoding of a qubit state for
  two particles: center spin-3/2 $u$ and the connected spin-1/2 $2'$ after a POVM on $u$ giving the outcome $a_u$.
  (b) This illustrates the case where the spin-3/2 particle is located at either end, and thus it has two neighboring
  qubits: $0$ and $2'$. (c) When the POVM's on two consecutive sites of center spin-3/2 particles give
  the same outcome, i.e., $a_u=a_v$, the qubit is encoded by the two blocks, consisting two spin-3/2's and two spin-1/2's.}
\end{figure}
Suppose the POVM on one of the center spin-3/2 $u$ results in
$a_u\in\{x,y,z\}$; see Fig.~\ref{fig:figEncoding}a. As seen from
Eq.~(\ref{POVM3}) or the stabilizer operators in
Eq.~(\ref{eqn:stabilizerF}), the three virtual qubits at $u$ are
projected to the subspace spanned by $|000\rangle$ and
$|111\rangle$, with $|0/1\rangle$ being the eigenstates of
$\sigma_{a_u}$ (and the quantization axis being $a_u$). Furthermore,
from the discussions above, there is a stabilizer operator for the
post-POVM state, as $-\sigma_{a_u}^{[2]}\otimes\sigma_{a_u}^{[2']}$
 commutes with $\tilde{F}_{u, a_u}$, where $2$ denotes the virtual
spin-1/2 on site $u$ that connects $2'$. Thus,
 the only two possibilities of the joint state of $u$ and the connected qubit $2'$ can be only
  $|(000)_u 1_{2'}\rangle$ and
$|(111)_u 0_{2'}\rangle$, as they satisfy
\begin{subequations}
\label{eqn:encoding1}
\begin{eqnarray}
&&-\sigma_{a_u}^{[2]}\otimes\sigma_{a_u}^{[2']}|(000)_u
1_{2'}\rangle=|(000)_u 1_{2'}\rangle, \\
&&-\sigma_{a_u}^{[2]}\otimes\sigma_{a_u}^{[2']}|(111)_u
0_{2'}\rangle=|(111)_u 0_{2'}\rangle.
\end{eqnarray}
\end{subequations}
 This
means that the two physical spins $u$ and $2'$ constitute the
encoding of an effective qubit.

\begin{table}
\begin{tabular}{l|c|c|c}
  \begin{tabular}{l} POVM \\ outcome\end{tabular} & {$z$} & {$x$} & {$y$} \\ \hline
    \begin{tabular}{l}Stabilizer \\ generators\end{tabular} &
    \begin{tabular}{c} $\sigma_z^{[1]}\sigma_z^{[2]}$, $\sigma_z^{[1]}\sigma_z^{[3]}$
     \\ and -$\sigma_z^{[2]}\sigma_z^{[2']}$\end{tabular} &
\begin{tabular}{c} $\sigma_x^{[1]}\sigma_x^{[2]}$, $\sigma_x^{[1]}\sigma_x^{[3]}$
     \\ and -$\sigma_x^{[2]}\sigma_x^{[2']}$\end{tabular} & \begin{tabular}{c} $\sigma_y^{[1]}\sigma_y^{[2]}$, $\sigma_y^{[1]}\sigma_y^{[3]}$
     \\ and -$\sigma_y^{[2]}\sigma_y^{[2']}$\end{tabular} \\ \hline
 \begin{tabular}{l}  Logical \\
 Pauli  ${X}$ \end{tabular} &  $\sigma_x^{[1]}\sigma_x^{[2]}\sigma_x^{[3]}\sigma_x^{[2']}$ &  $\sigma_z^{[1]}\sigma_z^{[2]}\sigma_z^{[3]}\sigma_z^{[2']}$&  $\sigma_z^{[1]}\sigma_z^{[2]}\sigma_z^{[3]}\sigma_z^{[2']}$
 \\ \hline
 \begin{tabular}{l}  Logical \\
  Pauli ${Z}$\end{tabular} & \begin{tabular}{c} $\sigma_z^{[1]},\sigma_z^{[2]}$,
 \\ $\sigma_z^{[3]}$ or  $-\sigma_z^{[2']}$\end{tabular}  & \begin{tabular}{c}$\sigma_x^{[1]}, \sigma_x^{[2]}$, \\
  $\sigma_x^{[3]}$ or
  $ -\sigma_x^{[2']}$\end{tabular}  &\begin{tabular}{c}$ \sigma_y^{[1]}, \sigma_y^{[2]}$,\\  $\sigma_y^{[3]}$ or  $-\sigma_y^{[2']}$\end{tabular}
  \end{tabular}
  \caption{\label{tbl:encoding}Table for local encoding. Please refer to Fig.~\ref{fig:figEncoding}a for the  qubit labeling \{1,2,3,2'\} at a site.
  The stabilizer generators determine the effective two levels, e.g., in the case of $z$-outcome: $|(000)1\rangle$ and $|(111)0\rangle$, where
 the qubits are listed in the order of labeling \{1,2,3,2'\}. The logical Pauli ${X}$ induces transition between the two, i.e., ${X}\sim |(000)1\rangle\langle(111)0|
 + |(111)0\rangle\langle(000)1|$. The logical Pauli ${Z}$ operators can differ by a sign due to the
 convention, e.g., one chooses ${Z}=\sigma_z^{1}$ and thus
 $ {Z}|(000)1\rangle=|(000)1\rangle$ and
 ${Z}|(111)0\rangle= - |(111)0\rangle$. If two consecutive
 sites share the same POVM outcome, e.g., $z$, then the encoding
 extends to two sites of total 8 qubits (6 virtual ones). If the
 site is at the end, as in Fig.~\ref{fig:figEncoding}b, a
 corresponding table of encoding using five qubits can be similarly constructed.
  }
  \end{table}

Now that we have identified the encoding of a qubit, we would like
to seek the encoding of Pauli X and Z operators. Let us suppose
$a_u=z$ and thus the two states $|\overline{0}\rangle\equiv|(000)_u
1_{2'}\rangle$ and $|\overline{1}\rangle\equiv|(111)_u
0_{2'}\rangle$ (with the basis $|0/1\rangle$ such that
$\sigma_z|0/1\rangle=\pm|0/1\rangle$) constitute the basis states of
a logical qubit. There are four equivalent choices of Pauli $Z$: (1)
$\sigma_z^{[1]}$, (2) $\sigma_z^{[2]}$, (3) $\sigma_z^{[3]}$ and (4)
$-\sigma_z^{[2']}$, as these satisfy that
$Z|\overline{0}\rangle=|\overline{0}\rangle$ and
$Z|\overline{1}\rangle=|\overline{1}\rangle$. The four choices are
equivalent, as any two of them are connected by a stabilizer
operator, e.g., $\sigma_z^{[2]}=\sigma_z^{[1]}(\sigma_z^{[1]}\otimes
\sigma_z^{[2]})$. To obtain the effective Pauli X, which has action
that takes $|\overline{0}\rangle$ to $|\overline{1}\rangle$ and vice
versa, we can check the following combination works,
\begin{equation}
\label{eqn:zX}
X\equiv\sigma_x^{[1]}\otimes\sigma_x^{[2]}\otimes\sigma_x^{[3]}\otimes\sigma_x^{[2']}.
\end{equation}

On the other, when $a_u=x$, the two basis states are
$|\overline{0}\rangle\equiv|(+++)_u (-)_{2'}\rangle$ and
$|\overline{1}\rangle\equiv|(---)_u (+)_{2'}\rangle$, where
$\sigma_x|\pm\rangle=\pm|\pm\rangle$. The four equivalent choices of
effective Pauli $Z$ are: (1) $\sigma_x^{[1]}$, (2) $\sigma_x^{[2]}$,
(3) $\sigma_x^{[3]}$ and (4) $-\sigma_x^{[2']}$, as these satisfy
that $Z|\overline{0}\rangle=|\overline{0}\rangle$ and
$Z|\overline{1}\rangle=|\overline{1}\rangle$. The effective Pauli X
can be chosen to be
\begin{equation}
\label{eqn:xX}
X\equiv\sigma_z^{[1]}\otimes\sigma_z^{[2]}\otimes\sigma_z^{[3]}\otimes\sigma_z^{[2']},
\end{equation}
as  $X|\overline{0}\rangle=|\overline{1}\rangle$ and
$X|\overline{1}\rangle=|\overline{0}\rangle$. Effectively, the POVM
outcome $a_u$ indicates the new quantization axis for the effective
qubit. For a summary of stabilizer generators, effective Pauli $X$
and $Z$ operators, please refer to Table~\ref{tbl:encoding}, which
also includes the case of $a_u=y$ outcome.

As exemplified in Fig.~\ref{fig:figEncoding}b, if the center
spin-3/2 is located at either end, it has two virtual spin-1/2's
connected to it. The three sites form the basis of the qubit
encoding: $|(000)_u 1_0 1_{2'}\rangle$ and $|(111)_u 0_0
0_{2'}\rangle$.

Now suppose the POVM's on several consecutive sites of center
spin-3/2 result in a same outcome; see the example in
Fig.~\ref{fig:figEncoding}c for two sites with $a=a_u=a_v$. Those
spin-3/2's and the connected virtual spin-1/2's together form the
encoding of a logical qubit. This is due to additionally that there
is a stabilizer operator $-\sigma_{a}^{[3]}\otimes \sigma_{a}^{[4]}$
for the virtual qubits on the two of the neighboring center sites
$u$ and $v$ with the same POVM outcome $a$ and that
$-\sigma_{a}^{[3]}\otimes \sigma_{a}^{[4]}$ commutes with both
$\tilde{F}_{u, a}$ and $\tilde{F}_{v, a}$. Therefore, the two blocks
that consists of two spin-3/2's ($u$ and $v$) and two spin-1/2's
($2'$ and $5'$) altogether only encode one qubit with two basis
states being $|(000)_u 1_{2'} (111)_v 0_{5'}\rangle$ and $|(111)_u
0_{2'}(000)_v 1_{5'}\rangle$. This result generalizes to any number
of consecutive center sites having the same POVM
outcome~\cite{Honeycomb}. When all spin-3/2 sites along a chain has
been performed of POVM, there is a sequence of POVM outcomes, e.g.,
$xxyzxzzzy\cdots.$ Consecutive sites of the same outcome (e.g., $xx$
and $zzz$) are grouped into what we call a domain.

We remark that by single-spin measurements, the encoding of a
logical qubit by a domain can be reduced to a single site. For
example, if one measures the spin at site $v$ in the basis
$|\pm\rangle\equiv (|0\rangle\pm|1\rangle)/\sqrt{2}$, a state
$\alpha|(000)_u 1_v\rangle +\beta|(111)_u 0_v\rangle$ will be
reduced to $\alpha|(000)_u\rangle \pm\beta|(111)_u \rangle$,
depending on the outcome of the measurement. Similarly, if one
measures the spin at site $u$ in the basis
$(|000\rangle\pm|111\rangle)/\sqrt{2}$, which is, in terms of
spin-3/2 language, $(|+3/2\rangle\pm|-3/2\rangle)/\sqrt{2}$, the
post-measurement state becomes $\alpha|0_v\rangle \pm\beta|1_v
\rangle$. Thus, an encoded cluster state can thus be reduced by
local measurement to a cluster state, where each site is effectively
a qubit. For simplicity, from now on we shall focus on cases where
no neighboring sites sharing the same outcome for illustration.

\subsection{Cluster-state stabilizer}
\begin{figure}
  \includegraphics[width=8cm]{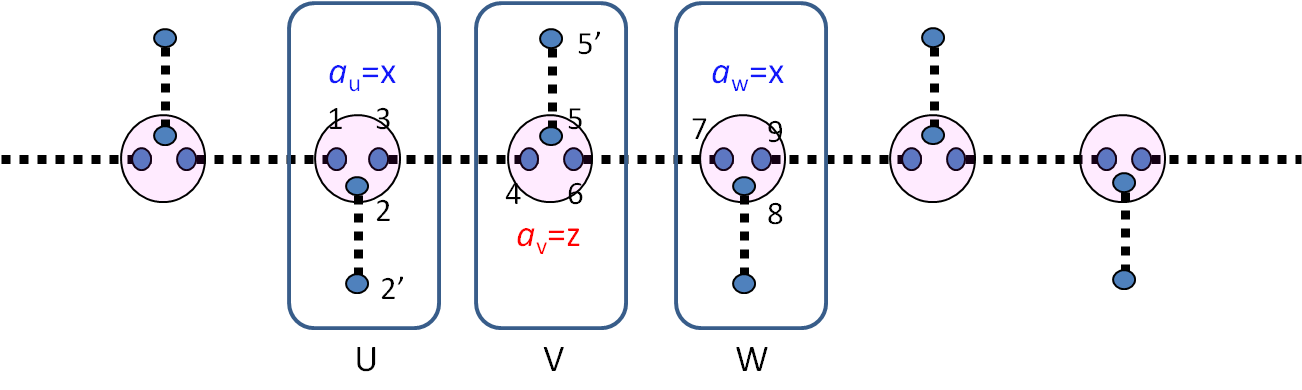}
  \caption{\label{fig:figZXZ}(Color
  online) Cluster-state stabilizer operator. The example shows POVM outcomes on the center sites $u$,
  $v$, and $w$ are $a_u=x$, $a_v=z$, and $a_w=x$, respectively. It can be shown that for the post-POVM state there is a stabilizer operator defined on
  the three blocks $U$, $V$, and $W$: $\overline{Z}_U \overline{X}_V \overline{Z}_W$, up to a possible sign.}
\end{figure}
To show that the post-POVM CMDB state is an (encoded) cluster state,
let us take the example shown in Fig.~\ref{fig:figZXZ}. Let us
labels the three center sites in blocks $U$, $V$ and $W$ by the
lower-case symbols $u$, $v$ and $w$ respectively. Suppose the POVM
outcomes on these sites are $a_u=x$, $a_v=z$, and $a_w=x$,
respectively. First note that $-\sigma_x^{[3]}\sigma_x^{[4]}$
commutes with $\tilde{F}_{u,x}$ and $-\sigma_x^{[6]}\sigma_x^{[7]}$
commutes with $\tilde{F}_{u,x}$. Note also that
$-\sigma_x^{[5]}\sigma_x^{[5']}$ is a stabilizer operator of the
singlet between $5$ and $5'$, but it does not commute with
$\tilde{F}_{v,z}$. However, if we multiply all the above operators,
we obtain \begin{equation} K_V\equiv- \sigma_x^{[3]}
\sigma_x^{[4]}\sigma_x^{[5]}\sigma_x^{[5']}\sigma_x^{[6]}
\sigma_x^{[7]}. \end{equation} Because $K_V$ commutes with
$\tilde{F}_{v,z}$, due to the identity
\begin{eqnarray}
&&\sigma_x\otimes\sigma_x\otimes\sigma_x\,
(|000\rangle\langle000|+|111\rangle\langle111|)\nonumber \\
& =&(|111\rangle\langle
000|+|000\rangle\langle111|)\nonumber\\
&=& (|000\rangle\langle000|+|111\rangle\langle111|)\,
\sigma_x\otimes\sigma_x\otimes\sigma_x,
\end{eqnarray}
it is a stabilizer operator of the post-measurement state. In terms
of logical Pauli operators ${Z}_U\equiv \sigma_x^{[3]}$ ,
${Z}_W\equiv \sigma_x^{[7]}$, ${X}_V\equiv
\sigma_x^{[4]}\sigma_x^{[5]} \sigma_x^{[6]} \sigma_x^{[5']}$ (see
the examples given in the previous section for the effective Pauli
operators), we arrive at the stabilizer operator $K_V=-{Z}_U {X}_V
{Z}_W$. This is (up to a sign) the stabilizer operator defining a
linear cluster state~\cite{Cluster}. We note that for completeness
of the qubit encoding for blocks $U$, $V$ and $W$, the corresponding
conjugate operators $Z$ or $X$  can be chosen as follows:
${X}_U\equiv \sigma_z^{[3]}$, ${X}_W\equiv \sigma_z^{[7]}$ and
${Z}_V\equiv \sigma_z^{[4]}$ and that they satisfy the required
anticommutation with their conjugate operators.

Note that the choice of three consecutive sites (or domains) is
arbitrary. Furthermore, if the above example is changed to a
four-site problem, e.g., with the POVM outcomes being $xzzx$, i.e.,
with middle two sites having the same $z$ outcome. These two sites
form a single logical qubit, as discussed earlier in
Sec.~\ref{sec:encoding}, and the encoded Pauli $X$ operators will
span across the two sites. The above example can be
straightforwardly generalized to a general proof that the state
after the POVM is, under local unitary transformation, equivalent to
an encoded cluster state, similar to the case of 2D AKLT state on
the honeycomb lattice~\cite{Honeycomb}. (The encoding of logical
qubits can be reduced to single sites by suitable local
measurements, as remarked earlier.) As a further illustration, let
us consider another example of three sites in Fig.~\ref{fig:figZXZ}
but with $a_u=x$, $a_v=z$, and $a_w=y$, i.e., the last site has a
different outcome $a_w=y$ than the above example. Because of this,
one now considers $-\sigma_y^{[6]}\sigma_y^{[7]}$ instead of
$-\sigma_x^{[6]}\sigma_x^{[7]}$ and can show that the follow
operator is a stabilizer generator:
\begin{equation}
 K_V\equiv- \sigma_x^{[3]}
\sigma_x^{[4]}\sigma_x^{[5]}\sigma_x^{[5']}\sigma_y^{[6]}
\sigma_y^{[7]}.
\end{equation}
Now, one can choose ${Z}_W\equiv \sigma_y^{[7]}$, ${Z}_U\equiv
\sigma_x^{[3]}$ , ${Z}_W\equiv \sigma_y^{[7]}$, ${X}_V\equiv
\sigma_x^{[4]}\sigma_x^{[5]} \sigma_x^{[6]} \sigma_x^{[5']}$, and
$Z_V\equiv \sigma_z^{[6]}$ (see Table~\ref{tbl:encoding}) and obtain
\begin{equation}
K_V=-{Z}_U ({X}_V i{Z}_V) {Z}_W=-{Z}_U {Y}_V {Z}_W.\end{equation}
 Although the stabilizer operator $K_V$ is not
of the canonical form of the cluster-state stabilizer ${Z}_U {X}_V
{Z}_W$ , they are related by local unitary transformation that
leaves $Z_V$ invariant.

The hidden cluster state in the ground state of the AKLT quasichain
is hence revealed by the POVM~(\ref{POVM1}) or
equivalently~(\ref{POVM2}). The original 1D spin-1 AKLT state has
previously been shown to be locally converted to a 1D cluster state
by a similar spin-1 POVM~\cite{Honeycomb} and by a different
alternating POVM~\cite{Chen10}.

\section{Quantum computational universality of the 2D Cai-Miyake-D\"ur-Briegel
state}\label{sec:2Dcluster}
\begin{figure}
   \includegraphics[width=8cm]{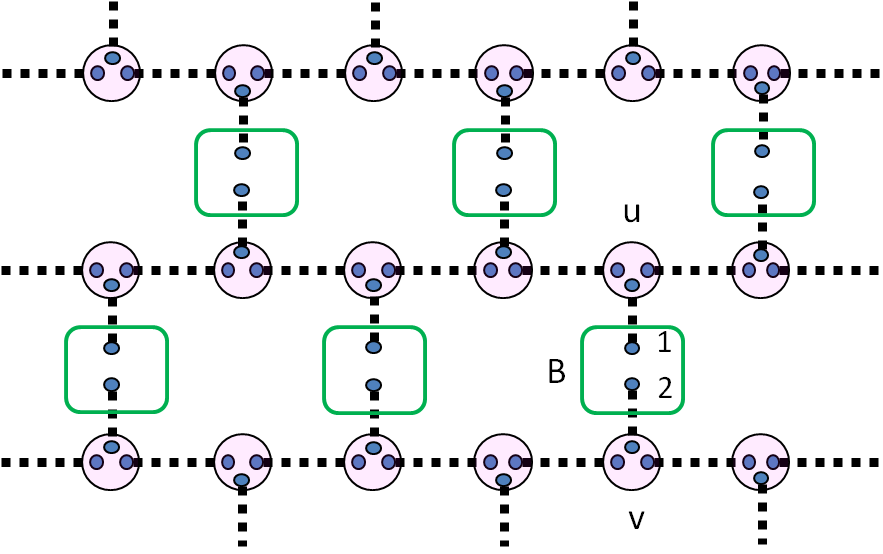}
  \caption{\label{fig:fig2D}(Color
  online) 2D structure. The two spin-1/2 $b$ particles (labeled $1$ and $2$) on two neighboring chains,
 are the two virtual qubits
 of the merged $B$-type spin-3/2 particle. By measuring the $B$ spin, two spins $u$ and $v$ will (i) evolves independently
 or (ii) be acted upon by a control-phase gate. }
\end{figure}

 After understanding that the quasichain
AKLT state can be converted to an (encoded) 1D cluster state,
arbitrary single-qubit rotations can be implemented along every
quasichain in the usual one-way computation linear cluster
states~\cite{Oneway}. What remains to be seen for the quantum
computational universality is whether entangling gates such as
control-NOT or control-phase can be implemented between the two
logical qubits on two neighboring quasichains. The answer is
affirmative.
 The two spin-1/2 $b$ particles on two neighboring chains,
 which are the two virtual qubits
 of the merged $B$-type spin-3/2 particle, are
used as either (1) a disconnecter or (2) a connecter so that either
in case (1)
 the two neighboring logical qubits evolve independently (without being acted upon by an entangling gate)
  or case (2)
an entangling gate acts on them.  Either functionality is achieved
by measurement on a $B$-site particle or equivalently by the joint
measurement of two $b$ particles; see Fig.~\ref{fig:fig2D}.
\subsection{Connecting/disconnecting neighboring logical qubits}
In the case of a disconnecter, the two virtual spin-1/2 particles
(of the single spin $B$) are each applied an effective Pauli X gate
$X=|0\rangle_a\langle 1|+|1\rangle_a\langle 0|$~\cite{applyX} before
being measured in the basis of respective $\ket{\pm}_a$, where the
axis label $a$ can be either $x$, $y$, or $z$ (depending on the POVM
outcome on the neighboring spin-3/2 particle) and $\ket{\pm}_a\equiv
(\ket{0}_a\pm\ket{1}_a)/\sqrt{2}$ with
$\sigma_a|0/1\rangle_a=\pm|0/1\rangle_a$; see Fig.~\ref{fig:fig2D}.
Suppose the outcome of the measurement on the spin-1/2 connected to
the spin-3/2 $u$ is labeled by $(-1)^{m_1}$ with $m_1=0/1$, and
similarly for the one connected to $v$: $(-1)^{m_2}$. (Together,
$m_1$ and $m_2$ characterize the measurement outcome of the merged
spin-3/2 $B$.) Then the effect of measuring the spin $B$ on
neighboring $u$ and $v$ is that single-qubit gates $Z_u^{m_1}\otimes
Z_v^{m_2}$ have been applied to $u$ and $v$, where
$Z_u=|3/2\rangle_{a_u}\langle 3/2|-|-3/2\rangle_{a_u}\langle -3/2|$
and similarly for $Z_v$. Note that in the three-virtual-qubit
picture, $|+3/2\rangle_{a_u}=|000\rangle_{a_u}$ and
$|-3/2\rangle_{a_u}=|111\rangle_{a_u}$. Furthermore, the measurement
on 1 and 2 corresponds to a local measurement on the spin-3/2 $B$.
The logical qubits on neighboring chains evolve independently, i.e.,
no entangling gate has been applied to $u$ and $v$.

In the case of a connecter, the two spin-1/2 particles are first
applied effective Pauli X gates: $X=|0\rangle_a\langle
1|+|1\rangle_a\langle 0|$ and then a control-phase (CP) gate: ${\rm
CP}_{12}= |0\rangle_{{a_u}}\langle 0|^{[1]}\otimes \openone^{[2]} +
|1\rangle_{a_u}\langle 1|^{[1]}\otimes Z_{a_v}^{[2]}$, where
$Z_{a_v}=|0\rangle_{a_v}\langle0|-|1\rangle_{a_v}\langle 1|$.  These
operations correspond a single-spin unitary transformation on the
merged $B$ spin. Then, a measurement in their respective
$|\pm\rangle_{a}$ basis is made as in the previous case, and the
effect is that an entangling gate has acted on the two associated
spin-3/2 particles $u$ and $v$ (which are effectively two qubits),
up to Pauli gates. Using the measurement labels $m_1$ and $m_2$, the
entangling gate that has been applied is $Z_u^{m_1}\otimes
Z_v^{m_2}\, {\rm CP}_{uv}$, where
\begin{equation}
{\rm CP}_{uv}= \Big\vert \frac{3}{2}\Big\rangle_{a_u} \Big\langle
\frac{3}{2}\Big\vert^{[u]}\otimes\, I^{[v]}_{a_v}\, +\, \Big\vert
-\frac{3}{2}\Big\rangle_{a_u} \Big\langle
-\frac{3}{2}\Big\vert^{[u]}\otimes \,Z^{[v]}_{a_v},
\end{equation}
where
$I^{[v]}_{a_v}=|3/2\rangle_{a_v}\langle3/2|+|-3/2\rangle_{a_v}\langle-3/2|$
is the effective qubit identity operator and
$Z_{a_v}=|3/2\rangle_{a_v}\langle3/2|-|-3/2\rangle_{a_v}\langle-3/2|$
is the effective Pauli Z operator. The logical qubits on neighboring
chains have thus undergone a control-phase gate up to Pauli Z's.

 We remark that the approach of
the joint measurement on the two virtual qubits has also been used
recently in Refs.~\cite{Cai10} and~\cite{Li} and was employed
earlier in Ref.~\cite{GottesmanChuang}. In summary, an entangling
gate can indeed be applied on two logical qubits residing on the two
neighboring quasichains. Thus, we see that CMDB state is a universal
resource for MBQC.
\begin{figure}
   \includegraphics[width=8cm]{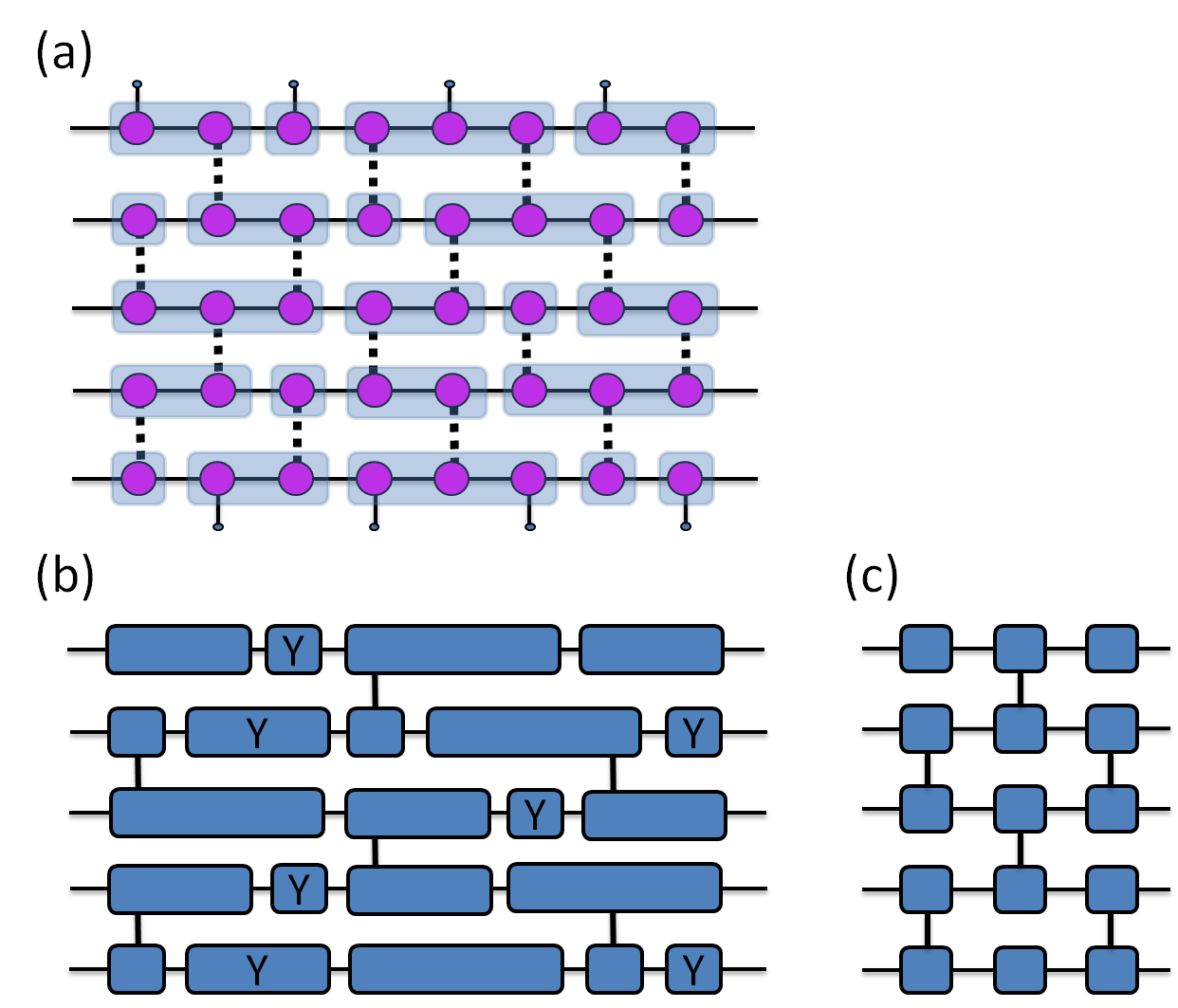}
  \caption{\label{fig:ToCluster}(Color
  online) Converting the Cai-Miyake-D\"ur-Briegel state to a 2D cluster state. (a) The shaded boxes enclose sites that share
the same POVM outcomes on $A$ sites. The $B$ sites are not shown
explicitly and only dashed lines are shown. (b) The solid vertical
lines denote the locations where CP gates are applied. Boxes labeled
with `Y' indicate subsequent logical Pauli Y measurements to be
made. (c) After the Pauli Y measurements, the left and right
neighboring logical qubits, represented by boxes, are linked. This
gives rise to a 2D cluster state. }
\end{figure}
\subsection{Converting to a 2D cluster state} The whole system can be
further converted to a 2D cluster state on a honeycomb lattice.
First, a structure which is topologically equivalent to a honeycomb
(or brick-wall) lattice can be identified. Second, by suitable
disconnecting or connecting measurements on $B$ sites and then Pauli
measurements on $A$ sites, the structure can be reduced to a
honeycomb lattice. Shown in Fig.~\ref{fig:ToCluster} is the example
of five quasichains. In Fig.~\ref{fig:ToCluster}a, the shaded boxes
enclose sites that share the same POVM outcomes on $A$ sites. The
$B$ sites are not shown explicitly and only vertical dashed lines
are used to indicate the possible connecter/disconnecter roles by
$B$ spins. In Fig.~\ref{fig:ToCluster}b, the solid vertical lines
denote the choice of a connecter (and the remaining dashed vertical
lines in Fig.~\ref{fig:ToCluster}a denote disconnecters) and hence a
control-phase gate acts on two neighboring $A$ sites connected by
solid vertical lines. The dangling spin-1/2's sites are omitted.
Those boxes with a `Y' represent a subsequent Pauli-Y measurement to
remove the qubit on the sites and to link its two neighboring
blocks, resulting in a graph state that has the brickwall structure,
as shown in Fig.~\ref{fig:ToCluster}c. This illustrates and
generalizes to a proof that a 2D cluster state can be distilled from
the CMDB state.

\section{Concluding remarks}\label{sec:conclusion}
In summary, we provide an alternative understanding of the quantum
computational universality of the 2D state constructed by Cai,
Miyake, D\"ur and Briegel~\cite{Cai10}. A crucial step is the
preprocessing given by the POVM~(\ref{POVM1}) on the central $A$
sites of all chains, which converts each chain to an encoded 1D
cluster state and thus arbitrary single-qubit rotation can be
implemented. Entangling gates, such as control-phase and control-NOT
gates, between neighboring two logical qubits  can be implemented by
measuring the linking spin-3/2 $B$ sites. The whole system can also
be further converted to a 2D cluster state.

The quantum computation in the correlation space proposed by Gross
and Eisert~\cite{Gross} was initially regarded as a scheme that
belongs to the so-called CC-universality~\cite{CCCQ}, where the
quantum computer takes classical input and delivers classical
output. Many of the resource states constructed in Ref.~\cite{Gross}
were later shown to be locally convertible to cluster
states~\cite{Cai09,Chen10}, thus enabling the so-called
CQ-universality, in which a quantum computer takes a classical input
but can deliver a quantum state as a output. Quantum computers of
CQ-universality seems potentially more powerful than those of
CC-universality. By showing that a 2D cluster state can be distilled
by local operations, we have in turn shown that the
Cai-Miyake-D\"ur-Briegel state enables a CQ-universal quantum
computation. Whether CQ-universality class is indeed more powerful
than CC-universality class in general remains open. In view of
resource states, this raises the question whether or not there
exists a resource state for which measurement-based quantum
computation can only be carried out in the correlation space but
cannot be done in the Hilbert space of physical spins.

\smallskip \noindent {\bf Acknowledgment.} We thank
Maarten Van den Nest for useful discussions. This work was supported
by NSERC (T.-C.W. and R.R.), MITACS (T.-C.W. and R.R.), CIFAR
(R.R.), the Sloan Foundation (R.R.), IARPA (R.R.) and  National
Research Foundation and Ministry of Education of Singapore
(L.-C.K.).


\begin{thebibliography}{99}
\bibitem{stabilizer2}
Stabilizer operators $K_v$ in the stabilizer ${\cal S}$ are unitary
operators that commute with one another. $K_v$ is a stabilizer
operator of the state $|\psi\rangle$ if it satisfies
$K_v|\psi\rangle=|\psi\rangle$. One important property of qubit
stabilizer operators is that when the number of independent
stabilizer operators equals that of the total number of qubits in
$|\psi\rangle$, the state is uniquely determined. If there is one
independent stabilizer less, then the state is only determined up to
a subspace of an effective qubit. For details, see
Ref.~\cite{Stabilizer}.
\bibitem{applyX}
The reason to apply a Pauli X is to flip the encoding states of the
$b$ and its assocaite $A$ particles from $|(000)_A 1_b\rangle$ and
$|(111)_A 0_b\rangle$ [see Eq.~(\ref{eqn:encoding1})] to $|(000)_A
0_b\rangle$ and $|(111)_A 1_b\rangle$, respectively, is for
convenience of later analysis. One then investigates the effect on
$A_1$ and $A_2$ due to the measurement on the associated $b_1$ and
$b_2$ in the $|\pm\rangle$ basis. In the case of a connecter, an
additional control-phase gate is applied to $b_1$ and $b_2$ before
measurement.
\end{thebibliography}
\end{document}